\documentclass[onecolumn]{emulateapj}

\usepackage{amsmath}
\usepackage{mathrsfs}
\usepackage{epsfig}
\usepackage{apjfonts}

\def\rd{{\rm d}}

\newcommand{\G}{\Gamma}

\newcommand{\e}{\epsilon}
\newcommand{\g}{\gamma}

\newcommand{\dD}{\delta_{\rm D}}

\newcommand{\psim}{\lower.5ex\hbox{$\; \buildrel \propto \over\sim \;$}}
\newcommand{\lbar}{\lower.0ex\hbox{$\; \buildrel {\lower0.0ex \hbox{-}} \over\lambda  \;$}}

\begin{document}

\title{A general relativistic external Compton-scattering model for TeV emission from M87}

\author{Yu-Dong Cui\altaffilmark{1}, Ye-Fei Yuan\altaffilmark{1}\footnote{corresponding author: yfyuan@ustc.edu.cn}, Yan-Rong Li\altaffilmark{2},
 \and Jian-Min Wang\altaffilmark{2}
 }
\affil{
1 Key Laboratory for Research in Galaxies and Cosmology, Department of Astronomy,
University of Sciences and Technology of China, CAS, Hefei, Anhui 230026, China\\
2 Key Laboratory for Particle Astrophysics, Institute of High Energy Physics, CAS,
19B Yuquan Road, Beijing 100049, China
}

\begin{abstract}
M87 is the first detected non-blazar extragalactic Tera-Electron-Volt (TeV) source with rapid variation and very flat spectrum
in the TeV band. To explain the two-peaks in the spectral energy distribution (SED) of the nucleus of M87 which is similar to
those of blazars, the most commonly adopted models are the synchrotron self-Compton scattering (SSC) models and the external
inverse Compton scattering (EIC) models. Considering that there is no correlated variation in the soft band (from radio to X-ray)
matching the TeV variation, and the TeV sources should not suffer from the $\g\g$ absorption due to the flat TeV spectrum, the
EIC models are advantageous in modeling the TeV emission from M87. In this paper, we propose a self-consistent EIC model to
explain the flat TeV spectrum of M87 within the framework of fully general relativity, where the background soft photons are
from the advection-dominated accretion flow (ADAF) around the central black hole, and the high energy electrons are from the
mini-jets which are powered by the magnetic reconnection in the main jet \citep{2010MNRAS.402.1649G}. In our model, both the TeV
flares observed in the years of 2005 and 2008 could be well explained: the $\g\g$ absorption for TeV photons is very low,
even inside the region very close to the black hole $20R_g\sim50R_g$; at the same region, the average EIC cooling time
($\sim 10^2\sim10^3s$) is short, which is consistent with the observed time scale of TeV variation.
 Furthermore, we also discuss the possibility that the accompanying X-ray flare in 2008
is due to the direct synchrotron radiation of the mini-jets.
\end{abstract}

\keywords{black hole physics --- galaxies: individual (M87)}

\section{Introduction}
Unlike blazars and the Galactic sources (e.g. SNRs), which composite the most population of TeV sources, M87 is the first
discovered radio galaxy with TeV radiation. Recently three other radio galaxies Centaurus A, 3C66B and IC310  were also
identified as TeV sources \citep{ 2009ApJ...695L..40A,2009ApJ...692L..29A, 2010ApJ...723L.207A}. Comparing to blazars, M87 has
much milder variations in both optical and X-ray band;
most importantly, the prominent kpc-scale jet of M87 has a very large viewing angle
$\sim30^\circ$ with respect to the line of sight \citep{bick96,ahar06}. In
2005, rapid TeV variation ($1-2$days) was discovered with flat
spectrum for the first time, but without correlated X-ray variation
from the nuclear core \citep{ahar06}. While in 2008, the radio, X-ray, TeV
joint observation discovered that TeV flares lasts 2 weeks
accompanied with an X-ray flare and a radio flare inside the
unresolved nuclear core ($30\times60R_s$) as well as with a radio blob moving out
of the unresolved core \citep{2009Sci...325..444A}. Here $R_s$ used
in the VLBA observations are based on the black hole (BH) mass of M87
$M_{BH}=6\times10^9M_{\sun}$ \citep{2009ApJ...700.1690G}. 
Recently, a TeV flare in 2010 which is very similar to the
previous TeV flares in 2005 and 2008 was reported. Further observations
at X-rays and radio find the correlated X-ray flare, but no
enhanced radio flux from the core \citep{Abramowski2011,Harris2011}.
In this paper, we adopt $M_{BH}=3\times10^9M_{\sun}$
\citep{1997ApJ...489..579M} to be consistent with the
advection-dominated accretion flow (ADAF) models and the mini-jet
models in the previous studies \citep{2009ApJ...699..513L,2010MNRAS.402.1649G}.

Although M87 has a large viewing angle, it is very near to
the Earth with a distance of $\sim16.7$Mpc. Owing to its
proximity, it is proposed that M87 might be a misaligned blazar
\citep{1998ApJ...493L..83T}.
Whereas blazars are believed to have jets beaming towards us, TeV sources in M87 could be shocks with relativistic
bulk velocity inside the jets. Recent observed minute-scale variation from galaxies like Mrk 421 (Fossati et al. 2008) and
PKS 2155-304 \citep{2007ApJ...664L..71A} indicates that TeV sources should be compact and very close to the BH. Comparing
to the TeV flares of blazars, the slower variation ($1-2$ day) detected in M87 could be the results of too few observed TeV
photons (each data point requires integration of photons for the whole night) or the much lower Doppler factor of the bulk
velocity of the TeV sources due to the large inclination. Furthermore, unlike the steeper TeV spectrum of blazars, the much
flatter TeV spectrum of M87 could be mainly due to the lack of $\g\g$ absorption. The large viewing angle may play an important
role in avoiding $\g\g$ absorption within the jet; thanks to its proximity and the dimness of its host galaxy, the absorption
to TeV photon from M87 caused by the galactic and the intergalactic background soft photons, as well as the cosmic
background photons is rather weak \citep{nero07}.

To explain the spectral energy distribution (SED) of the nucleus of M87, the one zone synchrotron self-Compton(SSC) models,
which have been applied to the TeV flares in blazars \citep[e.g.][]{1998ApJ...509..608T, 2001A&A...367..809K}, face
difficulties in fitting the two peaks in the SED, however, in the multi-zone SSC models, the TeV source is separated from
the soft ones, therefore the whole SED could be easily fitted as long as there are enough SSC blobs with specific locations,
electron distributions and bulk velocity \citep[e.g.][]{2008MNRAS.385L..98T,2007sf2a.conf..196L,2005ApJ...634L..33G}. The
multi-zone SSC models could also provide better explanation for the rapid variation, and the orphan TeV flares (the variations
in the soft band can not connect with the TeV variation). Furthermore, considering the very flat power-law SED of M87 in the TeV
band with index $p\simeq-2.2\sim-2.6$, the $\g\g$ absorption provides crucial constrains on the SSC models. So far, most of the
SSC models still have problems in explaining the very flat TeV spectrum due to the certain $\g\g$ absorption, therefore the
external inverse Compton (EIC) process could be more likely responsible for the TeV flare
\citep{2008MNRAS.384L..19B,2009MNRAS.395L..29G,2010MNRAS.402.1649G}. In the EIC models, the TeV source is far away from the soft
one, or has the relativistic bulk velocity with respect to the soft one, thus the $\g\g$ absorption can be reduced.

In those models mentioned above, how the very high energy (VHE) particles are produced is still an open question. 
It is generally believed that VHE particles are accelerated by the shocks in jets. 
There are some other possibilities include the mini-jets powered via the
magnetic reconnection in the main jet \citep[][and references therein]{2010MNRAS.402.1649G}; the magnetic centrifugal acceleration in the
vicinity of BH \citep{nero07,rieg08}, and so on. 
In the mini-jets model \citep{2010MNRAS.402.1649G}, magnetic reconnection within the main jet produces two oppositely directed (in the rest frame of the main jet) mini-jets, in the laboratory frame, one of them always points within the angle of the main jet and is observable in blazars because their jets point at us. The other mini-jet (its counterpart) points outside the opening angle of the main jet and is potentially observable to off-axis observers 
in case of the misaligned jets, such as those of M87 and Centaurus A.
In a word, the great advantage of the mini-jets model is that 
even at large inclination, 
mini-jets with a high bulk speed can still be detected, which is helpful for explanation of 
the fast TeV variation from M87.
After all, in order to avoid the $\g\g$ absorption, the energy density of the soft 
synchrotron photons in the TeV source must be limited, therefore, the minimal Lorentz factor of the VHE particles is generally assumed to be $10^{3-4}$ and the strength of the magnetic field in the TeV source below several Gauss \citep{2010MNRAS.402.1649G}.
Beside the direct inverse Compton process of the VHE electrons,
there are also some alternative models to produce the SED of the TeV
flares, such as the hadronic models including, among which, the interaction between the VHE protons and soft photons
\citep{2004A&A...419...89R} and the proton-proton collision process when a red giant was passing the base of the jet
\citep{2010ApJ...724.1517B}.

In this paper, we develop a fully general relativity EIC model for explaining the TeV emission from M87, in which the soft
photons emanate from the ADAF around the BH. In \S2, we investigate the safe zone of TeV photons (the $\g\g$ optical depth is
below unity). The technical details of our fully general relativity EIC model can be found in \S3. The numerical results and the
discussions are given in \S4 and \S5, respectively.

\begin{figure}[t!]
\centering
\includegraphics[width=0.6\textwidth, trim=0pt 80pt 0pt 0pt]{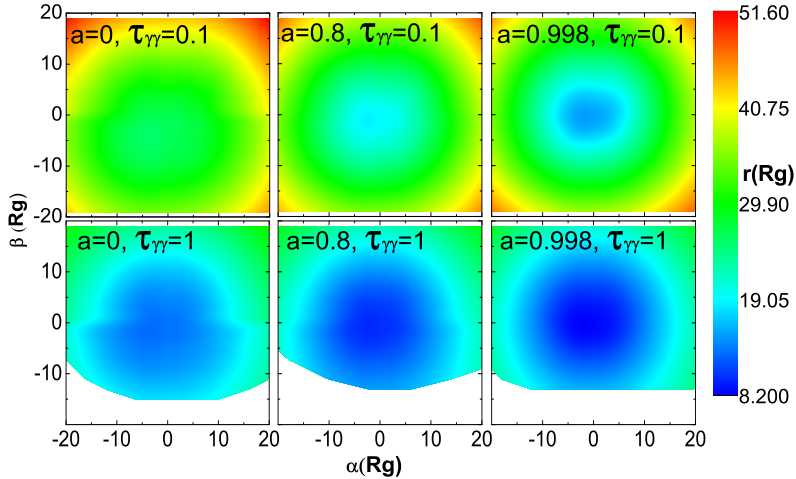}
\caption{ The contour of
$\tau_{\g\g}=0.1$ or $1$ in the $\alpha,\beta$ plane.
Where $r$ represents
the distance to the BH in the Boyer-Lindquist frame. The viewing angle
is taken to be $30^\circ$.  The blank region in the lower panels is the
region with $\tau_{\g\g} \leq 1$. }
\label{fig1}
\end{figure}

\section{Constraints on the location of TeV sources}
The rapid TeV variation of M87 ($t_{\rm var}<1-2$ days) found both in 2005,
2008 and 2010 indicates that the TeV photons should come from a compact source
very close to the black hole ($R_{\rm TeV}\lesssim ct_{\rm var}\delta_{\rm D}$,
where $\delta_{\rm D}$ is the Doppler factor of the source). Especially
the TeV flare in 2008, which is accompanied with a radio flare, lies inside
the unresolved region and there is a radio blob moving outwards.
Again, this indicates that the TeV
source might be near to the BH and its distance might be less than $\sim100R_{\rm s}$
\citep{2010HEAD...11.3006B}.

As mentioned above, beside the constrains from the VLBA radio image and the variation time scale of the TeV emission, the
background soft photons could also lay strong constrains on the location of the TeV sources, in consideration of the $\g\g$
absorption. Both the TeV flares in 2005, 2008 and 2010 had 
shown very flat power-law spectrum ($0.1\sim10$ TeV) strongly
suggesting quite weak $\g\g$ absorption. Therefore, the EIC process seems to be a more plausible mechanism to produce the flat
TeV spectrum, while the SSC models suffer from certain $\g\g$ absorption \citep{2010MNRAS.402.1649G}. Even in the EIC
models, if the source of soft photons is from a homogeneously isotropic blob whose size is about $50R_{\rm g}$ and
infrared luminosity is about $L_{\rm IR} \sim10^{41}$erg/s, the location of the TeV sources could be still limited to be no
deeper than $\sim 5R_{\rm g}$ inside that blob.

The observed soft photons from M87 could be mainly either from the disk or 
from the outflow/jet. According to the VLBI observations of M87, its jet 
is well known with a large inclination.
However, the base of the jet seems to have a large opening angle 
($\theta_{\rm open}\sim30^\circ$ inside the region
$\sim70R_{\rm s}$, \citealt{1999Natur.401..891J, 2007ApJ...660..200L,2009Sci...325..444A}),  therefore the jet might
contribute to the observed soft emission. Due to the slight variation of the observed flux of soft photons (from radio to X-ray)
from M87 nucleus, here we suggest that the observed soft photons are mainly from the accretion disk around the central black
hole as in Li et al. (2009).

By applying the ADAF model to the nuclear emission of M87,
Li et al. (2009) obtained the accretion rate of ADAF in M87, generally consistent
with the previous estimate of the Bondi accretion rate from the Chandra X-ray observation
(Di Matteo et al. 2003). They then calculated optical depth of the radiation fields from the ADAF
to TeV photons due to $\gamma\gamma$ absorption. The resultant optical depth suggests that
the location of TeV sources should be larger than $\sim10R_{\rm g}$,
in order to avoid the significant $\gamma\gamma$ absorption to a 10 TeV photon
\citep{2008ApJ...676L.109W,2009ApJ...699..513L}. 
Here we try to map a more detailed safe zone  ($\tau_{\g\g}\leq 1$) for
TeV photons in the vicinity of BH. As in \citet{2009ApJ...699..513L}.
For this purpose,
we calculate the optical depth ($\tau_{\g\g}$) to 10 TeV photons emanating
from vicinity of the central BH in the Kerr spacetime. The TeV photons move outwards
along their geodesic trajectories and reach the observer's sky at points
described by the impact parameters $\alpha$ and $\beta$.
Here $\alpha$ and $\beta$ respectively represent the displacement of the image
perpendicular to the projection of the rotation of the black hole
on the sky and the displacement parallel to the projection of
the axis (see e.g. Fig. 1 of Li et al. 2009).
The resultant contour of $\tau_{\g\g}$ as a function of the location of the TeV sources
for $\tau_{\g\g}=1$ and $0.1$ in the $\alpha\beta$ plane
can be obtained (see Fig.~\ref{fig1}).

As shown in Fig.~\ref{fig1}, it is obvious that the larger the spin $a$, the deeper the safe zone. For instance, for $a=0.998$,
the safe zone can reach even to $10R_{\rm g}$ along some trajectories of 10TeV photons, the reason is that for larger BH spin,
the infrared-UV radiation is concentrated in the inner disk. Therefore, the colliding angle between the soft photons and the TeV
photons in the mini-jets is smaller, significantly reducing the $\g\g$ absorption. Besides, in the panels of $a=0, 0.8$,
the asymmetricity is caused by the rotation of the disk and the viewing angle($30^\circ$) of the observer. While in the panels
of $a=0.998$, since the disk become a more compact source
of soft photons, the asymmetricity caused by the disk become weaker.

\section{Fully General Relativity External-Inverse Compton model}

\subsection{The TeV source}

In our EIC model, following \citet{2010MNRAS.402.1649G}, the mini-jets powered by magnetic energy
in the jet correspond to be the TeV sources. For a clarity, we present the details of the
model in what follows.

The strength of the magnetic field in the main jet is estimated as,
\begin{equation}
 {B_j^2\over 4\pi}=\left({L_{j,iso} \over 4\pi r_j^2 c \Gamma_j^2}\right)\left( {\sigma \over
 1+\sigma}\right),
\end{equation}
where $B_j$ is the magnetic strength, 
$L_{\rm j,iso}$ ($\sim10^{45}$ erg/s) is the observed isotropic power of the jet of
M87  \citep{bick96,reyn96,2000ApJ...543..611O,2006MNRAS.370..981S,2009ApJ...699.1274B},
$A_{\rm j}$ ($\pi {r_{\rm j}}^2$)
is the cross sectional area of the jet,  $\sigma$ is the ratio of the magnetization energy to the kinetic energy of the
main jet, $\Gamma_j$ is the Lorentz factor of the main jet, and $c$ is the light speed.

Although the inclination of the jet is about $30^0$, the mini-jets
(blobs of plasma with characteristic Lorentz factor $\Gamma_{\rm
co}$ at angle $\theta_{\rm mini}$ with respect to the jet in the jet
rest frame) are capable of beaming towards us with a high bulk
velocity
$\Gamma_{\rm em}=\Gamma_{\rm j}\Gamma_{\rm co}(1+\beta_{\rm j}\beta_{\rm co}\cos\theta_{\rm mini})$
in the laboratory frame. The bulk Lorentz factor of the mini-jets in
the jet rest frame is about $\Gamma_{\rm co}\sim \sigma^{1/2}$, which
corresponds to the Alfv\'{e}n speed of the plasma in the jet. The
characteristic thermal Lorenz factor of the electrons in the
mini-jets rest frame is about $\gamma'_{\rm ch}\sim
f\sigma^{1/2}m_{\rm p}/m_{\rm e}$. The energy distribution of the high energy
electrons is assumed to be
\begin{equation}
N^\prime_e(\g^\prime)=N_e(\g)/\delta_D^3=N_0 \gamma^{\prime p},
\end{equation}
where $10^4\lesssim\gamma^\prime=\g/\delta_D<\infty ,\ \
p\lesssim-3.2$. In our canonical model, $\Gamma_j$ is taken
to be 5, and $\Gamma_{\rm em}=12$ ($\delta_{\rm D}=23$, if the mini-jets
are beaming toward us). These parameters are
consistent with the original mini-jets model of \citet{2010MNRAS.402.1649G}.

\subsection{The Local Soft Radiation Field}
\begin{figure}
\centering
\includegraphics[angle=-90.0, width=0.6\textwidth]{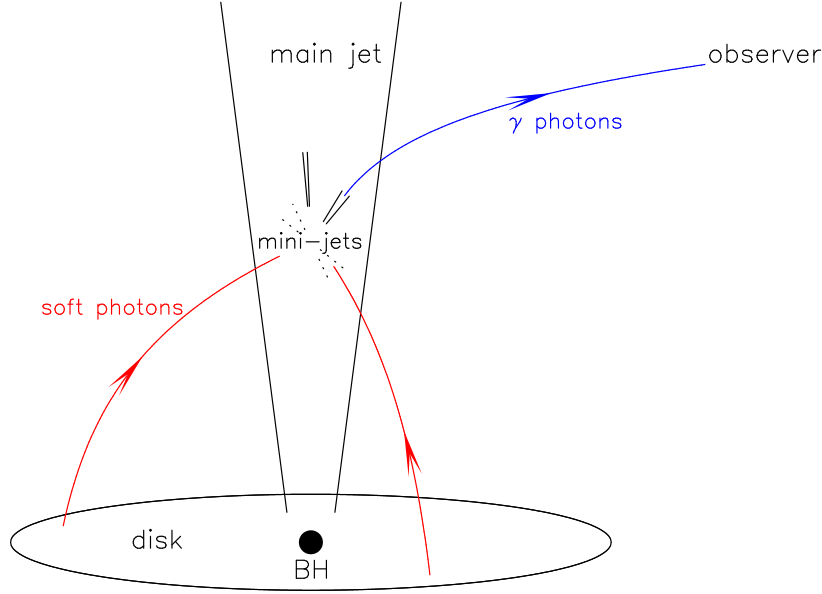}
\caption{A schematic picture for the disk dominating external
 Compton-scattering model. A mini-jet is beaming toward the infinity observer,
 and the soft photons are from the disk.}
\label{fig2}
\end{figure}

In our EIC model, we assume that the soft photons are from the accretion disk
and the TeV source (mini-jets) is located along the major axis of jet with
a height $H_{\rm TeV}$ above the central black hole.
To investigate the EIC process, we apply the same ray-tracing
technique as discussed in \citet{2009ApJ...699..513L}.
In \citet{2009ApJ...699..513L}, the authors discussed the optical depth
of the TeV photons due to colliding with the soft photons from the
accretion disk.
In this work, we investigate the Compton scattering of
the soft photons from the disk by the VHE electrons in the mini-jets
(as shown in Fig.~\ref{fig2}), therefore, we just simply replace the
TeV photons in the model of \citet{2009ApJ...699..513L} with the VHE electrons.

To obtain the flux density of the soft photons from the disk, the
global dynamical structure of the ADAF, such as the four velocity of
the fluid in the disk, the temperature of ions and electrons,
should be determined first; then the local emergent spectra
$I_{\nu_d}(r_{\rm d})$ at the radius $r_{\rm d}$ in the rest frame of the
fluid can be calculated. Using the ray-tracing technique, the
observed SED can be obtained to fit the multi-wavelength
observations of M87. The fitting parameters of the ADAF are obtained by
\cite{2009ApJ...699..513L}. In what follows, we summarize the
procedure to obtain the radiation energy density of the
soft photons from the direction ($\theta_{\rm s}, \phi_{\rm s}$)
 at the colliding location ($r_{\rm c},\theta_{\rm c},\phi_{\rm c}$).

\

In the locally non-rotating frame (LNRF), at the interacting place $(r_{\rm c},\theta_{\rm c},\phi_{\rm c})$ ,
the two motion constants of the soft photons ($\lambda_{\rm s}$, $\cal{Q}_{\rm s}$) are related with their traveling direction
$(\theta_{\rm s},\phi_{\rm s})$ as follows:
\begin{equation}
\lambda_{\rm s}=\frac{\mathscr{A}}{1+\omega\mathscr{A}},~~~ {\cal
Q}_{\rm s}=\mathscr{B}^2-(a\cos\theta_{\rm c})^2+(\lambda_{\rm
s}\cot\theta_{\rm c})^2.
\end{equation}
where,
\begin{equation}
\mathscr{A}=\frac{\sin\theta_s\sin\phi_s\sin\theta_{\rm c}
A}{\Sigma\Delta^{1/2}}, ~~~\mathscr{B}=\frac{\sin\theta_s\cos\phi_
sA^{1/2}(1-\omega\lambda_{\rm s})}{\Delta^{1/2}},
\end{equation}
where $A$, $\Sigma$, $\Delta$, and $\omega$ are the metric functions defined
in \citet{1972ApJ...178..347B}.

After knowing
the constants of motion $\lambda_{\rm s}$ and ${\cal Q}_{\rm s}$,
the soft photons can be traced back to the disk at certain radius
$r_{\rm d}$ by solving the geodesic equations (e.g.
\citealt{1972ApJ...178..347B,2009ApJ...699..722Y})
\begin{equation}
{\mathscr T}= \pm\int_{r_{\rm c}}^{r_{\rm d}}\frac{\rd r}{\sqrt{{\cal
R}(r)}}=\pm\int_{\theta_{\rm c}}^{\theta_{\rm d}}
              \frac{\rd\theta}{\sqrt{\Theta(\theta)}}
\end{equation}
where ${\mathscr T}$ is the affine parameter and the $\pm$ signs
represent the increment ($+$) or decrement ($-$) of $r$ and $\theta$
coordinates along the trajectory, respectively.

 Along these trajectories, the
redshift factor $g_{\rm s}$ for a soft photon travels from disk to the IC
interaction location can be obtained by
\begin{equation}
g_s = \frac{\nu_s}{\nu_d} =\frac{\left.e^\mu_{(t)}(\rm LNRF)P_\mu^{\rm
s}\right|_{r_{\rm c}}}
               {\left.e^\mu_{(t)}(\rm LRF)P_\mu^{\rm s}\right|_{r_{\rm d}}}
         =\frac{\left.e^{-\nu}(1-\omega\lambda_{\rm s})\right|_{r_c}}
           {\left.\gamma_r\gamma_\phi e^{-\nu}\left[1-\Omega\lambda_{\rm s}\mp
           \frac{\displaystyle \beta_r {\cal R}(r)^{1/2}}{\displaystyle  \gamma_\phi A^{1/2}}\right]
           \right|_{r_{\rm d}}}.
\end{equation}
where $\nu_{\rm s}$ and $\nu_{\rm d}$ are the frequency of the soft photon at the
colliding place and the disk, and
 $\gamma_r=(1-\beta_r^2)^{-1/2}$ and
$\gamma_\phi=(1-\beta_\phi^2)^{-1/2}$ are the Lorentz factors of the
radial and azimuthal velocity of the fluid in the accretion disk, respectively.

According to the Liouvell theorem, the final radiation energy density
of the soft photons at the IC location
($r_{\rm c},\theta_{\rm c},\phi_{\rm c}$) can be written as
\begin{equation}
U_{\rm s}(h\nu_{\rm s},\theta_{\rm s},\phi_{\rm s},r_{\rm c},\theta_{\rm c},\phi_{\rm c})=\frac{I_{\nu_{\rm d}}(h\nu_{\rm d},r_{\rm d})g_{\rm s}^3}{c},
\end{equation}
where $h$ is the Planck's constant.

\subsection{The VHE electrons}
After determining the radiation energy density of the soft photons
$U_{\rm s}$ at the colliding points, the direction of motion of the VHE electrons which produce the observed
TeV photons is needed to calculate the TeV spectra. Due
to the effects of special relativity, it is a {\rm reasonable} assumption that
both the direction of the relativistic electrons and that of the
TeV photons are the same. Denoting the two constants of the motion of
the TeV photons as $\lambda_{\rm IC}$ and $Q_{\rm IC}$, the beaming
direction of the TeV photons is as follows:

\begin{equation}
\cos\theta_{_{\rm IC}}
                    = \frac{\left.e^\mu_{(r)}P_{\mu}^{\rm IC}\right|_{r_{\rm c}}}
                          {-\left.e^\mu_{(t)}P_{\mu}^{\rm IC}\right|_{r_{\rm c}}}
                   =\frac{\pm {\cal R}(r)^{1/2}}{A^{1/2}(1-\omega\lambda_{\rm IC})},
\end{equation}

\begin{equation}
 \sin\theta_{_{\rm IC}}\cos\phi_{_{\rm
IC}}
                   =\frac{\left.e^\mu_{(\theta)}P_{\mu}^{\rm IC}\right|_{r_{\rm c}}}
                         {-\left.e^\mu_{(t)}P_{\mu}^{\rm IC}\right|_{r_{\rm c}}}
                   =\frac{\pm\Theta(\theta)^{1/2}\Delta^{1/2}}
                         {A^{1/2}(1-\omega\lambda_{\rm IC})},
\end{equation}
\begin{equation}
\sin\theta_{_{\rm IC}}\sin\phi_{_{\rm IC}}
                   =\frac{\left.e^\mu_{(\phi)}P_{\mu}^{\rm IC}\right|_{r_{\rm c}}}
                         {-\left.e^\mu_{(t)}P_{\mu}^{\rm IC}\right|_{r_{\rm c}}}
                   =\frac{\lambda_{\rm IC}}{\sin\theta_{\rm c}}
            \frac{\Sigma\Delta^{1/2}}{A(1-\omega\lambda_{\rm IC})},
\end{equation}
where the trajectory of the TeV photons determined by  $\lambda_{\rm IC}$ and $Q_{\rm IC}$
can be found by tracing the observed TeV photons from the infinity ($r=\infty,
\theta_{\rm obs} = 30^\circ)$ to the interacting location $(r=r_{\rm c},\theta=\theta_{\rm c})$.
As assumed above, the TeV source is located inside the
jet, therefore, $\theta_{\rm c}=0$ and subsequently $\lambda_{\rm IC}=0 $.

After determining the trajectory ($\lambda_{\rm IC},Q_{\rm IC}$) of the
TeV photons,
the corresponding redshift factor $g_{\rm IC}$ for the  $\gamma$-ray
photons traveling from the IC interaction location to the infinity is given by
\begin{equation}
g_{_{\rm IC}}=\frac{\nu_{\infty}^{\rm IC}}{\nu_{r_c}^{\rm IC}}
           =\frac{\left.e^\mu_{(t)}P_{\mu}^{\rm IC}\right|_\infty}
             {\left.e^\mu_{(t)}P_{\mu}^{\rm IC}\right|_{r_{\rm c}}}
           =\left.\frac{\Sigma^{1/2}\Delta^{1/2}}{A^{1/2}}\frac{1}{1-\omega\lambda_{\rm IC}}
            \right|_{r_{\rm c}}.
\end{equation}

\subsection{Spectrum of Compton-scattered external radiation fields}
Given the radiation energy of the soft photons and the energy
distribution of the high energy electrons, the Compton spectral
luminosity is given by (see \citealt{2008ApJ...686..181F}
and \citealt{2009ApJ...692...32D} for more details):

\begin{equation}
f_{local}^{\rm EC}(\e_{\rm IC}) = {\e_{\rm IC} L_{\rm
C}(\e_{\rm IC},\Omega_{\rm IC})\over d_L^2} = {c\pi r_e^2  \over 4\pi
d_L^2}\;\e_{\rm IC}^2\dD^3\;\int_0^{2\pi}d\phi_s \int_{-1}^1 d\mu_s\;
\int_0^{\e_{s,hi}} d\e_s\; {I_{\nu_{\rm d}}g_{\rm s}^3\over
\e_{\rm s}^2}\int_{\g_{\rm low}}^\infty d\g\; {N_e^\prime(\g^\prime )\over
\g^2}\; \Xi\;  ,
\end{equation}
where $\e_{\rm s}$ is the energy of the soft photons,
$\e_{\rm IC}$ is the energy of the $\g$-ray photons created via the IC
process, $\mu_{\rm s}=\cos(\theta_{\rm s})$, and $\mu_{\rm IC}=\cos(\theta_{\rm IC})$. With the
approximation that scattered photons travel in the same direction as
the VHE electrons, the Compton cross section can be drawn as (Dermer
et al. 1993 ; Dermer et al. 2006)
\begin{equation}
{d\sigma_{\rm IC}(\e_{\rm s},\e_{\rm IC},\gamma,\psi_{\rm IC}) \over d\e_{\rm IC}}
\;\cong\; {\pi r_e^2 \over \gamma\bar\e}\;\Xi\; \ ,\\ \  \left(
{\bar\e\over 2\gamma}< \e_{\rm IC} < {2\gamma\bar\e\over 1+2\bar\e} \right)\,
\end{equation}
\begin{equation}
\bar\e \equiv \gamma \e_{\rm s} (1-\sqrt{1-1/\g^2}\cos\psi_{IC})\cong
\gamma \e_{\rm s} (1-\cos\psi_{\rm IC}), \label{bare}
\end{equation}
 where
\begin{equation}
\Xi \;\equiv \; y+y^{-1}  - {2\e_{IC}\over \gamma \bar\e y} +
\left({\e_{IC}\over \gamma \bar\e y}\right)^2 \;, \label{Xi}
\end{equation}
\begin{equation}
y \;\equiv\; 1 - {\e_{IC}\over\g} \;, \label{y}
\end{equation}
and $\psi_{\rm IC}$ is the interaction angle between the VHE electron and the
soft photon during the IC process.
\begin{equation}
\cos\psi_{\rm IC} = \mu_{\rm IC} \mu_s +
\sqrt{1-\mu_{IC}^2}\sqrt{1-\mu_s^2}\cos(\phi_{IC}-\phi_s)\;,
\label{cospsi}
\end{equation}

\

The optical depth $\tau(\e_{\rm IC},H_{\rm TeV},a) $ of
these $\g$-ray photons ($\e_{\rm IC}$) can be integrated along their
trajectories as in Li et al. (2009),
\begin{equation}
\tau_{\g\g}(\lambda_{\rm IC},Q_{\rm IC},\e_{\rm IC})=\iiint
                               (1-\cos\psi_{\g\g})\sigma_{\gamma\gamma}(\e_s,\e_{\rm IC},\psi_{\g\g})
                   \frac{I_{\nu_{\rm d}}}{c\e_ s}g_{\rm s}^3\rd\Omega_s \rd\e_s \rd l,
\end{equation}
where  $\rd l=e^{\nu}\Sigma \rd\mathscr{T}$  is the proper length
differential with $\rd \mathscr{ T}$ defined to be differential of
the affine parameter $\mathscr{ T}$ along the trajectory of the TeV photons,
and $\psi_{\g\g}$ is the interacting angle
between the TeV photon ($\e_{\rm IC}$) and the soft photon ($\e_{\rm s}$).

Consequently, the final observed SED in the infinity would be,
\begin{equation}
f_{\rm Earth}^{\rm EC}\left(\e_{\rm IC}{g_{\rm IC}\over 1+z}\right) =
e^{-\tau(\e_{\rm IC})}\left({g_{\rm IC}\over 1+z}\right)^4 f_{local}^{\rm
EC}(\e_{\rm IC}).
\end{equation}

\section{Numerical results}
\subsection{$ H_{\rm TeV}, a$ dependence}
In this paper, we choose the BH spin as $a=0,\ 0.8,\ 0.998$
to be consistent with the ADAF model used in fitting the SED (from
radio to X-ray) of M87  in \cite{2009ApJ...699..513L}. The location of
the TeV source is set to be very close to the BH as $H_{\rm TeV}=5, 10,
20, 50$, and $100R_g$, respectively.
The distribution of the VHE electrons is described
by $N_0=0.4\times10^{50}$, $\gamma^{'}_{\rm min}=10^4$, and $p=-3.2$,
where the maximal index $p$
is chosen to fit the very flat TeV flare spectrum observed in 2005
and 2008 \citep{2010MNRAS.402.1649G}.

\begin{figure}
\centering
\includegraphics[width=0.8\textwidth]{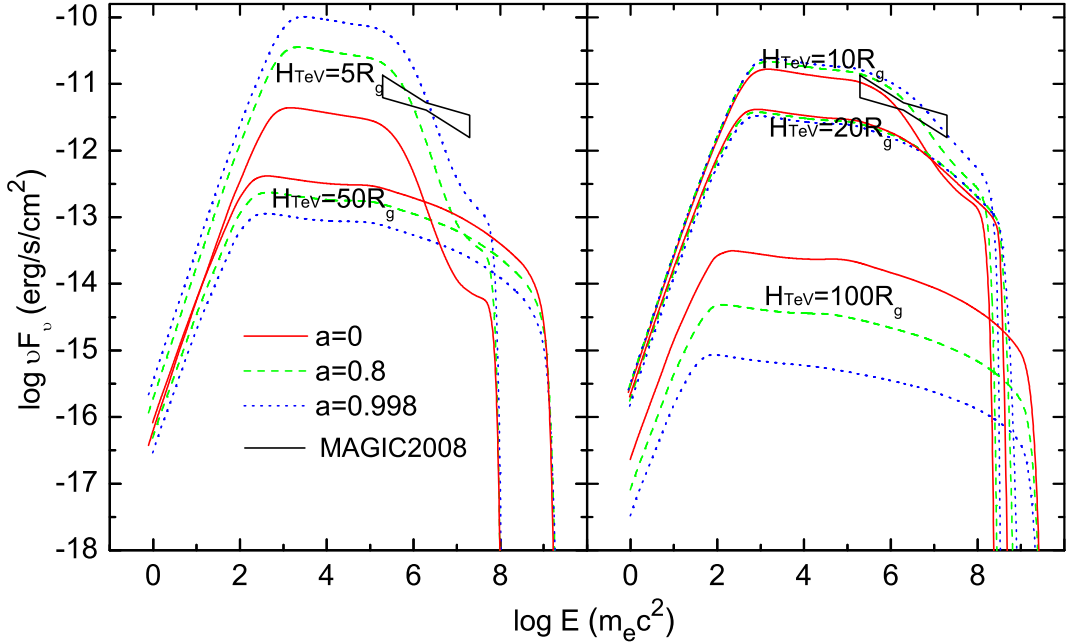}

\caption{SED of the external
Compton-scattering model with the different spins of the black hole ($a=0,0.8,0.998$) and the location of the mini-jets is above
the black hole ({\it Left panel:} $H_{\rm TeV}=5,50 R_{\rm g}$, {\it right panel:} $H_{\rm TeV}=10,20,100 R_{\rm g}$). The
energy distribution of the VHE electrons are determined by three parameters: $N_0=0.4\times10^{50}$, $\gamma_{\rm min}=10^4$ and
$p=-3.2$. The Lorentz factor of the mini-jets in the laboratory frame is taken to be $\Gamma_{\rm em}=12$.  MAGIC observation in
2008 is shown in the symbol of butterfly (Albert et al. 2008).} \label{fig3}
\end{figure}

In Fig.~\ref{fig3}, we show the dependence of the SED of the
EIC scattering on the height of the TeV source and BH spins. We can find that
the efficiency of the
Compton-scattering increases dramatically if the TeV source is closer
to the disk. Furthermore, if the location of the TeV source $H_{\rm TeV}$ is
below $20R_{\rm g}$, the Compton-scattered luminosity of the soft
radiation fields from the the disk around the black hole with spin
$a=0.998$ is higher than that from the black hole with spin $a=0$;
meanwhile, if the location of the TeV source is higher than $20R_{\rm g}$,
for $a=0$, the Compton-scattered luminosity is higher than that for
$a=0.998$.
This is because for $a=0.998$, the soft radiation field from the
disk is more compact, i.e., the soft radiation field is strong in
the vicinity of the BH, however, at the distance far away from
the disk, the region of the soft radiation field looks like a point
source. Therefore, the colliding angle between the soft photons and
the VHE electrons in the mini-jets is smaller, which reduces the
rate of the inverse Compton scattering significantly.

It is also clearly shown in Fig.~\ref{fig3} that each spectrum has
an exponential cut off beyond
$E_{\rm cut}\sim10^{8\sim9}m_{\rm e}c^2$ which is caused by the $\g\g$
absorption. If the location of the TeV source is nearer to the BH
($H_{\rm TeV}<20R_{\rm g}$), the spectrum will also suffer from the severe $\g\g$
absorption in the observable band $0.1-10$TeV. In each of those
reduced spectrums of $H_{\rm TeV}=5, 10$, there is always a warp up
trend at $E_{\rm IC}\sim10^{8} m_{\rm e}c^2$ which is caused by the
relatively low optical depth $\tau_{\g\g}$ at that band, since the
spectrum of the disk has a relatively low luminosity at
$E_{\rm s}\sim10^{-2}m_{\rm e}c^2$.

\subsection{TeV flares and The X-ray companion}
In 2008, observations had shown that the TeV, X-ray, and radio flares
are well correlated \citep{2009Sci...325..444A},
though no X-ray correlation was found
in 2005.  We propose that the accompanying X-ray flare in 2008 results from
the synchrotron radiation of the VHE electrons in the mini-jets, in other words,
the X-ray flare could be a byproduct of
the TeV flare, which just happens to be above the background X-ray
flux. While in 2005, it is more likely the X-ray flaring flux is
below the more stable X-ray background which is attributed to the
disk and/or the corona. It is notified that the TeV flare in 2008 is
stronger than that in 2005, and the corresponding X-ray flare is
rather mild: during the flaring, the X-ray flux only increased about
two times above the average.

\begin{figure}
\centering
\includegraphics[width=0.8\textwidth]{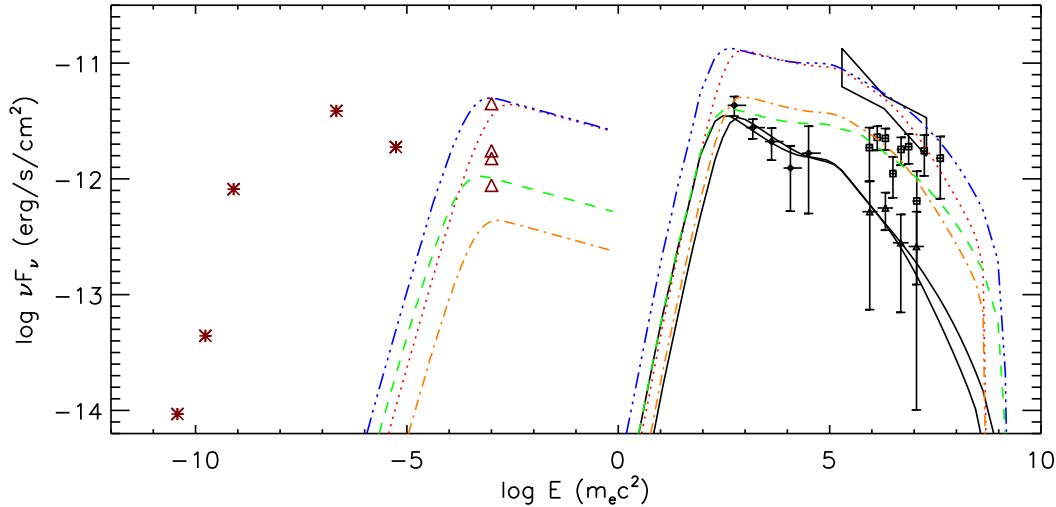}
\caption{ Synchrotron (dashed lines and
dash-dotted lines) and the external Compton-scattered (solid lines
and dotted lines) emission from a Mini-jet which are modeled to fit
the observed emission from M87, the detailed model parameters are
shown in Table one. The observed emission include: $(1)$. MAGIC
observation in 2008 which is shown in the symbol of butterfly
(Albert et al. 2008); $(2)$. HESS observation in 2005 which is
indicated by the black square
   with long error-bars ;
$(3)$. HESS observation in 2004 which is labeled as the black triangle
   with long error-bar (Aharonian et al. 2006);
   $(4)$. FERMI observation in 2009 which is labeled as the black
   solid circle with error-bars (Abdo et al.2009);
$(5)$. The approximate data from Chandra is shown in the hollow triangles. With 5\% uncertainty, the one on the top
represent the flux of nucleus in 2008 Feb 16 (data are provided by Dr. D. E. Harris), the other three represent the flux
in 2005 Apr 22, Apr 28, May 04 separately \citep{2006ApJ...640..211H,2009ApJ...699..305H};  $(6)$. Typical nucleus emission is
shown in the asterisk.\citep{harm94, maoz05,reyn96,spar96}.
 The synchrotron-self Compton emission is too weak to
be shown in this figure.} \label{fig4}
\end{figure}

\begin{deluxetable}{cccccc}
\tabletypesize{\footnotesize}
\tablecolumns{6}
\tablewidth{0.8\textwidth}
\tablecaption{Model parameters used in fitting the observations
shown in Fig.4.} \tablehead{ \colhead{EIC lines in Fig.4} &
\colhead{$H_{\rm TeV}(R_{\rm g}$)} & \colhead{$\g_{\rm min}$} & \colhead{
$N_0$(counts)} & \colhead{$p$} & \colhead{$B(G)$} }

\startdata

\sidehead{ \hspace*{0.2\textwidth}Fitting 2008MAGIC\tablenotemark{a}
with two lines on the top}
 dash-3doted line    &50  &$10^4$   &$1.0 \times10^{51}$           &-3.2  &2.6 \\
dotted line    &20    &$10^4$   &$1.0 \times10^{50}$           &-3.2  &8 \\
\hline
 \sidehead{ \hspace*{0.2\textwidth} Fitting 2005HESS\tablenotemark{b} with two lines in the middle}
   dashed line   &50  &$10^4$   &$0.6 \times 10^{51}$  &-3.2 &1.6  \\
 dash-dotted line  &20    &$10^4$   &$0.4 \times 10^{50}$  &-3.2  &4 \\
\hline
 \sidehead{ \hspace*{0.2\textwidth}Fitting 2004HESS\tablenotemark{c}+2009FERMI\tablenotemark{d} with two lines at the
bottom}
solid line &50  &$10^4$     &$1.6\times10^{52}$ &-3.5  &-- \\
solid line &20  &$10^4$  &$0.5\times10^{51}$  &-3.5 &--  \\
\enddata
\tablecomments{The bulk speed of the mini-jets in local lab frame
is taken as $\Gamma_{\rm mini}=12$ and the viewing angle is set as
$\theta_{\rm obs}=30^\circ $. The parameter spin of the
BH is chosen as $a=0.8$ , because the changing of the spin barely
influence the shape of the spectrum, it just slightly boost the
entire SED up or down when applying $H_{\rm TeV}=20,50R_g$.}
\tablenotetext{a}{MAGIC observation of M87 in 2008, as a flaring state}
\tablenotetext{b}{HESS observation of M87 in 2005, as a flaring state}
\tablenotetext{c}{HESS observation of M87 in 2004, as a quiet state}
\tablenotetext{d}{FERMI observation of M87 in 2009, as a quiet state }

\end{deluxetable}

In Fig.~\ref{fig4}, both the SED of the direct synchrotron radiation and the Compton-scattered radiation are shown. We choose
$H_{\rm TeV}=20, 50$ to be as close to the BH as possible while avoid severe $\g\g$ absorption. Owing to the weak $\g\g$
absorption, all the EIC spectra have
 inherited the original power-law index of the VHE electrons, which is $p=-3.2,
 -3.5$. As expected, the corresponding synchrotron
 spectrum fitting the X-ray flares in 2008
  is comparable with or exceeds the average X-ray flux, while the X-ray flux in 2005
 is buried beneath the background flux.

The magnetic field in the mini-jets is taken to be $B=1.6, 4, 2.6, 8$G, respectively, which is used to calculate the direct
synchrotron radiation. The strength of the magnetic field used in
this work is consistent with that in the original mini-jet model
$B_{\rm em}$ \citep[][and references therein]{2010MNRAS.402.1649G}, that
is, $\lesssim 0.8 (L_{\rm j,iso}/10^{45}$ erg/s)$^{1/2} ({100 R_{\rm g}} /
r_{\rm jet}) (5/\Gamma_j)^2$ G.
  We propose that the base of jet
is cone shaped as $r_{\rm jet}\sim 0.5 H sin\theta_{\rm open}$, where the
full opening angle $\theta_{\rm open}\sim30^\circ$
\citep{2009Sci...325..444A}, which leads to $B(50R_{\rm g}) \lesssim 6.4$G,
$B(20R_{\rm g}) \lesssim 16$G. The more detailed parameters are shown in
Table 1.

There are four main parameters $N_0, p, H_{\rm TeV}, a$ in our models.
Where the spin $a$ is fixed because it only slightly influence the
SED for $H_{\rm TeV}=20,50$ as shown in Fig.~\ref{fig3}; the electron
power-law index $p$ is fixed to follow the observed TeV spectrum
index and the $\g\g$ absorption is too low to soften the spectrum;
the other two parameters $N_0$ and $H_{\rm TeV}$ are degenerated, making
it hard to draw any solid constrains by fitting the SED. To be more
specific, the final TeV flux should be roughly $F_{\rm IC}\sim
N_0/H_{\rm TeV}^2 $. As shown in Table 1, to fit the same observation
data, when the $H_{\rm TeV}$ changes from $20R_{\rm g}$ to $50R_{\rm g}$,
we need
about 10 times more VHE electrons. Therefore, considering that both the
TeV flares in 2005 and 2008 have the same $a$ and $p(-2.2\sim-2.6 )$,
we suggest that the TeV source in 2008 may be more powerful or/and be
closer to the BH.

Being closer to the BH, the TeV source will suffer more $\g\g$
absorption, which would soften the spectrum. But no obvious changes
of the spectrum index can be drawn from the observations in 2005 and
2008 due to the long error-bars. It is noticed that, as shown in
Fig. 3, the $\g\g$ absorption barely influences the $\g$-ray spectrum
below 0.1 TeV, therefore the Fermi telescope should be able to
provide better constrains on the power-law index of the VHE
electrons.

As shown in Fig.~\ref{fig4}, we also fit the quiet
state (2004HESS+2009FERMI) spectrum of M87 with the mini-jets model.
In other words, the quiet TeV flux could be attributed from some
mini-jets which are weaker, misaligned, or heavily absorbed by
synchrotron or/and $\g\g$ pair production (lower $N_{\rm e,p},\delta_D$).
Where the misaligned Mini-jets could be considered as weaker
mini-jets consisted of much fewer VHE electrons beaming towards us
with lower bulk velocity, see details in
\citep{2010MNRAS.402.1649G}. Therefor we could not exclude the
possibility that weaker flares are due to misaligned mini-jets,
which are supposed to have lower $\delta_{\rm D}$. It should be noticed
that lower $\delta_{\rm D}$ could not explain the steeper spectrum.

As mentioned above, the soft photons field are demanded to satisfy
both the low $\g\g$ absorption rate and the high IC cooling
efficiency. In models as shown in Fig.~\ref{fig4}, apparently, the
synchrotron cooling effect is not dominating since the EIC flux is
much higher. The IC cooling time is short enough so that there is  no influence on
the observed rapid variation ($1\sim2$days):
\begin{equation}
{ E_{\rm e} \sim L_{\rm GeV-TeV}t_{\rm cool}/(4\Gamma_{em}^2) ,
\ \ E_{\rm e}\sim
\G_{\rm em}\int_{10^4}^{\infty} \g^\prime N^\prime_e(\g^\prime) d
\g^\prime , \ \ t_{\rm cool}<t_{\rm var}},
\end{equation}
where $L_{\rm GeV-TeV}$ is the isotropic total EIC flux and $E_{\rm e}$ is 
the total energy of VHE electrons in the lab
frame. With the parameters listed in Table 1, as expected, 
one can find $t_{\rm cool}\sim 10^{2-3}$s. 
During this cooling time, the mini-jets can
only travel less than $\sim 0.5R_g$, which is consistent with 
the above assumption that the location of the TeV source 
keeps unchanged during a flare event.

\section{Conclusions and Discussions}

In this paper, by taking into account the fully general relativity effects, we propose a disk-dominating external
Compton-scattering model for explaining the flat TeV radiation from M87. The external Compton-scattering model suffers much less
self-$\g\g$ absorption, comparing to the one-zone self-synchrotron Compton model, in which the soft photons and the VHE
electrons are from the same blobs. The advantage of the EIC model in which the soft radiation is from the accretion disk is that
the soft radiation source could be more compact and further away from the TeV source than ones in the SSC models, in which the
$\g\g$ absorption can be reduced significantly. Besides, our EIC model is also supported by the observed mild IR-UV-X flux
variations of the nucleus and the large viewing angle of the jet of M87 (the nuclei flux with viewing angle $30^\circ$ is
unlikely to be contaminated by the jet, neither to be blocked by the disk itself). 
It should be noticed that 
the VHE electrons and TeV photons may be influenced by some outflows
with the anisotropic radiation at the soft band which are not beaming towards us, 
such as the jet itself. Unfortunately, the base of the jet of M87 appears 
to be quite chaotic, thus it is hard to draw any conclusions about the 
detailed structure of jet/outflows \citep{2011arXiv1109.6252P}.
In our model, the TeV sources are the mini-jets which are easily able to beam 
towards us with high Doppler factor
and give rise to the observed VHE radiation.

How to distinguish between SSC and EIC models by the future observations?
The main difference between these two models is that
the IR-UV variation and TeV flares are not necessarily correlated in the 
EIC models, but they are in the SSC models, especially the multi-zone SSC model.
Therefore, we can distinguish between the SSC and EIC models by the 
simultaneous observations of the IR-UV emission during the TeV flare. 
Unfortunately, during the 2005 TeV flares, there is no accompanied HST 
observation of M87. With the better sensitivity of CTA (it is about $>$10 
times better than those of the present Cerenkov
telescopes), we will be able to detect the fainter flares and obtain the 
minimal variation time scale of the TeV flares, which increases the 
opportunity of finding the correlated variation between IR-UV and TeV emission.

In our models, with little $\g\g$ absorption, all the EIC spectra at the observational band have inherited the original
power-law index of the VHE electrons. After fitting those spectra, we calculate the cooling time of the VHE electrons in the
mini-jets (see section 4) to make sure that they would not travel longer than $ct_{\rm var}\delta_D$. It turns out
when TeV source is located at $H_{\rm TeV}<50R_{\rm g}$, the IC cooling time is very much shorter than the upper-limits of the
observed TeV variation of 1-2 days. Therefore, if possibly, the TeV source 
is located further away from the
disk ($H_{\rm TeV}>50R_{\rm g}$), the cooling efficiency of the VHE electrons in the mini-jets is much lower, the mini-jets
could move out of the main jet. As a result, the variation of the soft radiation background at the different locations above
the disk should be considered.

We also discuss the probability that the direct synchrotron radiation
from the
mini-jets may cause the correlated X-ray flare in 2008. As expected,
according to our model, if the magnetic field is about several Gauss,
the direct
synchrotron flux from these mini-jets which lies at the X-ray band
can explain the X-ray flare very well but has no influence on the
observed radiation at the GeV-TeV band.
However, the magnetic field $B$ can not be
constrained by any direct observation, as discussed above,
and we only have a rough estimate of the strength of the magnetic
field supposed to be
consistent with the mini-jets model. Considering that the synchrotron
flux is roughly about $F_{\rm syn}\sim N_0B^2$, it is obvious that the
$N_0$ and $B$ are degenerated. If the TeV source is
confined inside the region $20R_g<H_{\rm TeV}<50R_{\rm g}$,
the corresponding strength of the magnetic field is required to be
about $2G< B < 8G$, in order to produce the X-ray flare in 2008.

Dealing with the two weeks data of the 2008 flare from MAGIC, Albert et al. (2008) had separated the high state data ( $\sim$ 1
day time scale) from the low state data and they found that during the high states, the spectra are harder:
the spectrum index of the high state is about 2.2, while that of the low state is about 2.6.
Because the IC cooling time and the synchrotron cooling time
of VHE particles is short, 
the observed fastest variation should be dominated by the time of the acceleration,
or even the shifting of the beaming angle of the mini-jets.
If there is no continuous acceleration of the relativistic electrons, 
the spectra of the TeV flares will become steeper.
In the year of 2008, VERITAS captured the TeV flare (Feb 9-13, 2008) which is 
close to the date of the X-ray flare (Feb 16, 2008). VERITAS observation
shows that the spectrum index of the TeV flare
is about 2.4, which is softer than the earlier flares captured by MAGIC
\citep{2010ApJ...716..819A}.
We suggest that the TeV flare captured by VERITAS 
is happening during the rapid synchrotron cooling
of the relativistic electrons in the mini-jet, and the X-ray flare 
could be a byproduct of that TeV flare, 
as the X-ray flares in 2008. Our suggestion
could naturally explain the X-ray excess and the softer 
spectrum of TeV emission.
There are two ways to check above suggestion: 
first, the flux of the X-ray via synchrotron radiation 
should be anti-correlated with the TeV flux via EIC,
which could be tested by the future observations with the higher time
resolution (less than 1 day); 
second, and the TeV flare should share the similar spectrum index with 
that of the X-ray flare. Unfortunately, due to the pile up effect of 
X-ray observation \citep{2009ApJ...699..305H}, we can not obtain 
the correct spectrum index of the X-ray flare.

\acknowledgements
We would like to thank the anonymous referee for her/his constructive 
suggestions and comments, and 
Dr. Daniel Harris for providing his preliminary results
on {\it Chandra} observations of M87. This work is partially supported by
National Basic Research Program of China (2009CB824800),
the National Natural Science Foundation (11073020,10733010,11133005),
and the Fundamental Research Funds for the Central Universities (WK2030220004).

\end{document}